\documentclass[]{aastex631}

\shorttitle{Further analysis for galactic dark matter halos with pressure}
\shortauthors{Ace\~na et al.}

\graphicspath{{./}{figures/}}

\begin{document}

\title{Preliminary Study on Dark Matter-Dominated Systems: Further Analysis for galactic dark matter halos with Pressure}

\author{A. Ace\~na}
\altaffiliation{acena.andres@conicet.gov.ar}
\affiliation{Instituto Interdisciplinario de Ciencias B\'asicas, CONICET, Facultad de Ciencias Exactas y Naturales \\
Universidad Nacional de Cuyo \\
Mendoza, Argentina}

\author{J. Barranco}
\altaffiliation{jbarranc@fisica.ugto.mx}
\affiliation{Departamento de F\'isica, Divisi\'on de Ciencias e Ingenier\'ias \\
Campus Le\'on, Universidad de Guanajuato \\
Le\'on 37150, M\'exico}


\author{A. Bernal}
\altaffiliation{bernal.a@ugto.mx}
\affiliation{Departamento de F\'isica, Divisi\'on de Ciencias e Ingenier\'ias \\
Campus Le\'on, Universidad de Guanajuato \\
Le\'on 37150, M\'exico}

\author{E. L\'opez}
\altaffiliation{ericsson.lopez@epn.edu.ec}
\affiliation{Observatorio Astron\'omico de Quito y Departamento de F\'isica de la Facultad de Ciencias, \\
Escuela Polit\'ecnica Nacional \\
Quito, Ecuador}

\author{M. Llerena}
\affiliation{Universidad de la Serena, Chile}

\begin{abstract}
Low surface brightness galaxies are an excellent laboratory where stars and baryonic matter act as tracers of the gravitational potential of the dark matter halo. If dark matter is modeled as a perfect fluid, then spherically symmetric and static dark matter halos in hydrostatic equilibrium demand that dark matter should have an intrinsic pressure that counteracts the gravitational attraction that the dark matter halo exerts on itself.  This static fluid (dark matter-dominated system, where the presence of baryons is negligible) has a specific equation of state for each rotational velocity profile of the stars in galaxies. In this work, we study the dark matter equation of state needed for the self-gravitating object to produce a gravitational potential such that the tracers follow the  Universal Rotational Velocity Profile for stars of spiral galaxies proposed by Persic et al. (1996) and analyze the properties of the self-gravitating structures that emerge from this equation of state.The resulting configurations explaining the observed rotational speeds are found to be unstable. We conclude that the halo is not in hydrostatic equilibrium, it is nonspherically symmetric, or it is not static if the universal velocity profile should be valid for fitting the rotational velocity curve of the galaxies.  
\end{abstract}

\keywords{Galaxies: general, haloes --- Cosmology: dark matter}

\section{Introduction} 

Current astrophysical observations at cosmological and galactic scales suggest a concordance standard model coined Lambda Cold Dark Matter ($\Lambda$-CDM) \citep{Planck:2018vyg}. It contains three major components: a cosmological constant $\Lambda$, a cold dark matter component (CDM), and ordinary matter. 
We don't know much about this invisible component named CDM that acts gravitationally on baryonic matter, being the observed rotational curves of spiral galaxies one of the most direct evidence of its existence \citep{Sofue:2000jx}.

It is not known what it is made of but there is the belief that it is a pressureless medium that dominates in the outer regions of spiral galaxies. While luminous matter dominates in the innermost regions of galaxies, it appears that the effects of dark matter can also be found in regions where ordinary matter is present \citep{Persic1996}.
The pressureless condition of CDM leads to a background cosmological evolution of its density that varies as $a^{-3}$,  being $a$ the scale factor of the Universe. Latest satellite missions WMAP and Planck have promoted cosmology to a new precision era where the hypothesis of pressureless dark matter can be tested \citep{Muller:2004yb,Serra:2011jh,Calabrese:2009zza,Xu:2013mqe,Yang:2015rta,Kopp:2018zxp}. In particular, it has been found that a barotropic equation of state for the dark matter, $p_{DM}=\omega_{DM}~\rho$, with $\omega_{DM}=0.000707^{+0.000747}_{-0.000746}$ is compatible with the evolution of the Universe  \citep{Xu:2013mqe,Yang:2015rta,Kopp:2018zxp}, hence 
the hypothesis of dark matter with a small pressure can not be ruled out in favor of the pressureless CDM hypothesis.

Nevertheless, at galactic scales, it is known that $\Lambda$-CDM is unable to provide a complete description of the dark matter halos \citep{Weinberg:2013aya,Perivolaropoulos:2021jda}. The core-cusp problem \citep{Flores:1994gz,Karukes:2015fma}, the too big to fail problem \citep{Boylan-Kolchin:2011qkt} and the missing satellite problem \citep{Klypin:1999uc,Moore:1999nt}, among others, force us to carefully reconsider the pressureless dark matter hypothesis. It is known that self-interacting dark matter could solve some small-scale problems of $\Lambda$-CDM \citep{Spergel:1999mh}. The strength of self-interaction between dark matter particles leads to some effective pressure. Moreover, dark matter must have pressure to avoid intermediate-mass black holes increasing their mass far beyond observations due to dark matter accretion  \citep{Pepe:2011dk,Lora-Clavijo:2014kha}.
While the Lambda Cold Dark Matter ($\Lambda$-CDM) model has been successful at explaining a wide range of observational data at large scales ($\Lambda$-CDM assumes that dark matter particles are cold and non-relativistic), it faces some challenges and discrepancies at smaller scales, particularly on galaxy scales. Among alternative dark matter models that currently aim to address the galaxy-scale issues of the $\Lambda$-CDM model, we have Self-Interacting Dark Matter (SIDM) \citep{Spergel:1999mh}: In the standard $\Lambda$-CDM model, dark matter is considered to be collisionless, meaning that dark matter particles do not interact with each other except through gravity. SIDM proposes that dark matter particles can have self-interactions, leading to a more efficient transfer of angular momentum during galactic collisions. This could potentially address discrepancies between simulations based on $\Lambda$-CDM and observed galaxy properties. Warm Dark Matter (WDM) \citep{Bulbul:2014}: Warm dark matter models propose that dark matter particles have some non-negligible thermal velocities, falling between the cold and hot dark matter scenarios. WDM can suppress the formation of small-scale structures, potentially addressing the "missing satellites" problem observed in certain galaxy surveys. Fuzzy Dark Matter (FDM) \citep{Hu:2000pu}: FDM introduces a scalar field as a candidate for dark matter. The de Broglie wavelength associated with the scalar field introduces a characteristic length scale, which can potentially resolve issues related to the core-cusp problem in the density profiles of dwarf galaxies. FDM could lead to a "fuzziness" in the density distributions of dark matter on small scales. Modified Newtonian Dynamics (MOND) \citep{Milgrom:1983}: MOND is an alternative approach that modifies the laws of gravity rather than introducing new types of dark matter particles. It suggests that at very low accelerations, gravity does not follow the standard Newtonian behavior. MOND has been successful at explaining galactic rotation curves without the need for dark matter, but it faces challenges in explaining observations at larger scales. Axion Dark Matter \citep{Sikivie:1983}: Axions are hypothetical elementary particles that were originally proposed to solve the strong CP problem in particle physics. They are also considered candidates for dark matter. Axion dark matter can form a Bose-Einstein condensate on galactic scales, leading to unique observational signatures.
It is important to note that while these alternative models offer interesting possibilities, none of them have been definitively proven, and the nature of dark matter remains an open question in astrophysics and particle physics.\\

On the other hand, the equation of state (EOS) of dark matter in virialized structures has been a topic of significant interest and research within the astrophysics and cosmology communities. Understanding the EOS is crucial for determining the properties and behavior of dark matter on galactic and larger scales. Several efforts have been made to establish the EOS and uncover the physical processes underlying it: observational data, such as galaxy rotation curves, cosmic microwave background (CMB) radiation, and large-scale structure surveys, provide important constraints on the distribution and behavior of dark matter \citep[e.g.]{Navarro:1996}. N-body simulations, hydrodynamic simulations, and semi-analytical models simulate dark matter structure formation, offering insights into its EOS by modeling gravitational collapse and virialization of dark matter halos \cite[e.g.]{Springel:2005}. The cold dark matter (CDM) paradigm and alternative theories (e.g., warm dark matter, self-interacting dark matter) describe dark matter properties and its EOS, validated against observations and simulations. Studies focus on determining fundamental particle properties (e.g., mass, interaction cross-section) influencing dark matter's EOS in virialized structures. Feedback from star formation, supernovae, and active galactic nuclei impacts dark matter distribution and behavior, crucial for understanding its EOS \citep[e.g.]{Hopkins:2018}. The community's efforts to establish the EOS of dark matter in virialized structures involve a combination of observational, theoretical, and numerical approaches. In the case of dark matter halos, the polytropic EOS is often used to describe the pressure-density relationship of the dark matter component. However, it's important to note that the exact form of the EOS for dark matter halos can vary depending on the specific model or simulation being used.  

In the context of dark matter in virialized structures, the equation of state (EOS) typically refers to the relationship between the pressure $p$ and the energy density $\rho$ of dark matter. For non-relativistic matter like CDM, the pressure is typically negligible compared to the energy density, so the EOS is often assumed to be $p\sim 0$. This simplification arises because the speed of dark matter particles in virialized structures is much smaller than the speed of light ($v \ll c$), so the kinetic energy of the particles is much smaller than their rest mass energy, and thus the pressure can be neglected in many cases.  It is worth noting that the exact nature of dark matter and its EOS are still open questions in astrophysics and cosmology, and different theoretical models may propose different EOS depending on the specific properties assigned to dark matter particles.\\

This junction, between the actual observational capability for testing the pressureless hypothesis of dark matter and the current problems of 
$\Lambda$-CDM to explain some issues at galactic scales, is a strong motivation to study the possibility that dark matter has some intrinsic pressure.  
It is so as by considering the rotational curves of galaxies in \cite{Barranco:2013wy}, it was shown that given a velocity profile $v_t$ for test particles in a  
spherically symmetric and static dark matter halo in hydrostatic equilibrium,   the gravitational potential $\Phi$ is fixed by $v_t$ 
trough $\frac{d\Phi}{dr}=\frac{v_t^2}{r}$. If the dark matter in the halo is modeled as a perfect fluid, two variables describe the fluid: the mass density $\rho(r)$ and the pressure $p(r)$. Einstein's equation in spherical symmetry, once the gravitational potential is known (fixed by the rotational curve profile $v_t(r)$), relates $p(r), \rho(r)$ and the mass. That is, once the gravitational potential is fixed, then the hydrostatic equilibrium equations automatically determine an effective equation of state, $p(\rho) \ne 0$, for dark matter given $v_t(r)$.
That is to say, to determine the equation of state, it considers different distributions of dark matter density and rotation curves and solves the Tolman Oppenheimer Volkov (TOV) equations. The free parameters of the equation of state are fitted to best match the observed rotation curves of the selected galaxy groups. In this way, several equations of state (EoS) were obtained in \cite{Barranco:2013wy}. Each EoS corresponds to a different velocity profile.\\

The present work explores in more detail one of those EoS by analyzing the structure of the resulting dark matter halos obtained as self-gravitating structures of a perfect fluid that models dark matter with such EoS. In particular, the EoS that we are going to explore is the one that is obtained using the universal rotational velocity profile studied by Persic and Salucci in \cite{Persic1996}. In this case, the EoS is given by \cite{Barranco:2013wy}:
\begin{equation}
    \rho(p) = \frac{\rho_\bullet}{6}\left(1+8\frac{p}{p_\bullet}-\sqrt{1+8\frac{p}{p_\bullet}}\right), \label{EOS_rho}
\end{equation}
where $\rho_\bullet$ and $p_\bullet$ are free parameters. If one considers a particular galaxy and fits the universal rotational velocity profile, then this fitting provides two particular values to $\rho_\bullet$ and $p_\bullet$. In the present work we perform what can be considered the inverse procedure to what was done in \cite{Barranco:2013wy}. We start with the EoS (\ref{EOS_rho}) with the parameters $\rho_\bullet$ and $p_\bullet$ as given, and make the ansatz that this is the true EoS for dark matter. Then, we solve the Tolman-Oppenheimer-Volkov equations with such EoS, obtaining the matter profile $m(r)$. For this, we need to prescribe the central pressure as the initial data in the ODE system, and then numerically integrate the Tolman-Oppenheimer-Volkov equations. To each central pressure, there corresponds one particular distribution of dark matter, that we consider as a possible dark matter halo. By varying the central pressure we obtain the family of dark matter halos, all with the same EoS, given by (\ref{EOS_rho}). Once the matter profile $m(r)$ for one such object is known, the velocity profile $v_t$ for that object is straightforward to obtain. In the family of objects obtained, only one will have a velocity profile in the form of the universal rotational velocity profile, this object will be the one for which the central pressure is $p_\bullet$. In our analysis, for a given set of values for the two parameters, $\rho_\bullet$ and $p_\bullet$, the resulting dark matter halos show several deficiencies to be astrophysical reasonable  dark matter halos, namely:
\begin{itemize}
\item The radial density decays as $\rho(r)\sim 1/r^2$ and extends to infinity. 
\item The total mass of the resulting dark matter halo grows linearly in $r$.
\item The resulting self-gravitating configurations that produce rotational curves that fit the observational data are in the branch of unstable configurations. 
\end{itemize}
The first two shortcomings can be resolved by defining the dark matter halo radius where the halo density equals the average density of the universe. The problem of the instability of the halo implies that at least one of our assumptions is incorrect: either the halo is not in hydrostatic equilibrium, or it is not spherically symmetric, or it is not static, or dark matter is not a perfect fluid, or the universal rotational velocity profile does not provide an accurate fit to the velocity profile, especially far away from the galactic center.\\

The present work consists of a
preliminary study on dark matter-dominated systems, i.e., the presence of baryons is negligible. Furthermore, we assume a system in virial equilibrium and propose an equation of state (EOS) for dark matter halos. It is crucial to understand that this EOS specifically applies to the galactic scale. This limitation arises because the EOS is derived from the rotation curves of spiral galaxies and is not intended for use in a cosmological context.
Virialization occurs when a system of gravitationally interacting particles has evolved to the point where the kinetic energy of its constituents, such as particles, galaxies, or dark matter, is roughly balanced by the gravitational potential energy. This balance leads to a stable configuration.  
One approach is to study the EOS of dark matter in the context of specific dark matter models, such as cold dark matter (CDM) or warm dark matter (WDM). These models make different predictions about the behavior of dark matter particles, including their velocities and interactions, which can affect the EOS in virialized structures. Furthermore, theoretical efforts aim to incorporate the effects of dark matter self-interactions, if present, into models of virialized structures. These interactions could affect the EOS and lead to observable differences in the distribution of dark matter compared to models without self-interactions.
Overall, theoretical advancements play a crucial role in understanding the EOS of dark matter in virialized structures, providing insights that complement observational and numerical studies.\\

To show how these conclusions are attained, the article is organized as follows: Section \ref{section_2} reviews the TOV equations and explains the EoS (equation \ref{EOS_rho}) analyzed in this work. Section \ref{section_3} details the numerical methods here employed and shows the resulting self-gravitating structures obtained, as well as, their main properties. Section \ref{section_4} is devoted to some discussions and then Section \ref{section_5} is for the conclusions regarding the obtained results.

\section{Self-gravitating perfect fluid as a dark matter halo}\label{section_2}
Dark matter will be treated as a perfect fluid defined by stresses $T^r_r=T^\theta_\theta=T^\phi_\phi=p$, $T^t_t=-\rho$, being $p$, $\rho$ the pressure and the density of the fluid, respectively. The rest of the stresses are zero. This can be seen as the simplest possible departure from CDM. Galaxies are composed of luminous matter encapsulated by a dark matter halo. Observation of carbon giant stars in the Galactic halo implies that the dark matter halo of our galaxy is spherically symmetric \citep{Ibata:2000pu}. Since our galaxy is not special, it seems reasonable to assume that most dark matter halos are spherically symmetric and thus we will assume such symmetry. 
As the amount of luminous matter compared with the amount of dark matter in Low Surface Brightness (LSB) galaxies is small, then baryonic matter does not contribute significantly to the total mass of these galaxies. As such, 
the halo for LSB galaxies can be modeled by a self-gravitating sphere of perfect dark matter fluid in hydrostatic equilibrium. Therefore, in \cite{Barranco:2013wy}, it is assumed that the baryonic matter in the galaxy acts only as test particles that probe the gravitational potential. The gravitational potential itself is due to the dark matter content of the galaxy. For the dark matter halo not to collapse, there must be an intrinsic pressure that the dark matter impinges upon itself. As the dark matter does not interact with the baryonic matter, this pressure does not affect in any way the trajectories of the baryonic test particles. Dark matter interacts only by gravitation, that is
say that its behavior is purely non-collisional; however, the behavior of the baryons is given by the collisional dynamics dissipating energy through
radiative atomic processes; radiative cooling. Then, in this work dark matter with an intrinsic pressure is considered. The pressure produced, for instance, by massive neutrino free-streaming may contrast the gravitational collapse and maintain the halo equilibrium. Considering the equilibrium conditions of dark matter, the simplest assumption is to consider the dark matter-like-ideal gas of non-relativistic particles with a density $\rho (r)$ and pressure $p(r)$, which are related by a state equation. The meaning of dark matter pressure is not clear as well. However, the DM particle velocity distribution is the pressure analogy that provides the halo stability against the gravitational field. 
In this section, we present the equations that will describe such a self-gravitating structure:
the Tolman-Oppenheimer-Volkov equations and the dark matter EoS.

\subsection{Tolman-Oppenheimer-Volkov equations}
First, we recall the well-known general relativistic Tolman-Oppenheimer-Volkov
(TOV) equations \citep{Tolman1939,OppenheimerVolkov1939} (see
also \cite{Silbar2004}), which are the main theoretical tool used in the present work. We consider a static and spherically symmetric spacetime, whose line element in Schwarzschild coordinates is
\begin{equation}
    ds^2 = -e^{2\Phi}c^2dt^2+\frac{dr^2}{(1-\frac{2Gm}{c^2r})}+r^2(d\theta^2+\sin^2\theta \,d\phi^2),
\end{equation}
being $m$ and $\Phi$ functions of only the radial coordinate $r$. We think of $m$ as the gravitational mass inside the sphere of radius $r$ and $\Phi$ can be interpreted as the Newtonian gravitational potential. If the matter content of the spacetime is a perfect fluid, then the Einstein field equations imply the TOV system of equations:
\begin{equation}
    \frac{dm}{dr} = 4\pi r^2\rho, \label{TOVmass}
\end{equation}
\begin{equation}
    \frac{dp}{dr} = -\frac{Gm\rho}{r^2}\frac{\left(1+\frac{p}{c^2\rho}\right)\left(1+\frac{4\pi r^3 p}{mc^2}\right)}{1-\frac{2Gm}{c^2r}}
    \label{TOVpressure}.
\end{equation} 
These equations express the equilibrium at each $r$, between the internal pressure that the material supports against the attraction of the gravitational mass within $r$. These are the hydrostatic equilibrium equations in General Relativity, where the factor $\frac{2G m}{c^2r}$, that appears in the equation (\ref{TOVpressure}), determines whether the effects of General Relativity should be taken into account or not. 
When  $\frac{2G m}{c^2r}\ll1$ we can neglect this term in the TOV system and we arrive at the corresponding well-known Newtonian limit: 
\begin{equation}
    \frac{d m}{d r} = 4\pi r^2\rho, \quad
    \frac{dp}{dr} = -\frac{G m \rho}{r^2}.
\end{equation}

It is convenient for the discussion and numerical integration to use dimensionless quantities,
\begin{equation}
    \bar{m} = \frac{m}{M_\star},\quad \bar{r} = \frac{r}{R_\star},\quad \bar{\rho} = \frac{\rho}{\rho_\star} ,\quad \bar{p} = \frac{p}{p_\star},
\end{equation}
where $M_\star$ and $R_\star$ are the characteristic scales for mass and distance of the system under study. In \cite{Barranco:2013wy}, they are taken to be
\begin{equation}
    M_\star = 10^{10} M_\odot,\qquad R_\star = 1\,kpc.
\end{equation}
Also, the characteristic density and pressure are
\begin{equation}
    \rho_\star = \frac{M_\star}{\frac{4}{3}\pi R_\star^3},\qquad p_\star =c^2 \rho_\star,
\end{equation}
which give the values
\begin{equation}
    \rho_\star = 1.66 \times 10^{-22}\frac{g}{cm^3},\qquad p_\star = 1.49\times 10^{-1}\frac{g}{cm\,s^2}.
\end{equation}
In terms of these quantities equations \ref{TOVmass}-\ref{TOVpressure} take the form
\begin{equation}\label{TOV1}
    \frac{d\bar{m}}{d\bar{r}} = 3\bar{r}^2\bar{\rho},
\end{equation}
\begin{equation}\label{TOV2}
    \frac{d\bar{p}}{d\bar{r}} = -G_\star\frac{\bar{m}\bar{\rho}}{\bar{r}^2}\left(1+\frac{\bar{p}}{\bar{\rho}}\right)\left(1+\frac{3\bar{r}^3\bar{p}}{\bar{m}}\right)\left(1-2G_\star\frac{\bar{m}}{\bar{r}}\right)^{-1},
\end{equation}
with
\begin{equation}
    G_\star = \frac{GM_\star}{c^2R_\star} = 4.785\times 10^{-7}.
\end{equation}

The system of equations \ref{TOV1}-\ref{TOV2} is formally singular at $\bar{r}= 0$,  being $\bar{p}(\bar{r}=0) = \bar{p}_0$ the only free parameter. To perform the numerical integration of these equations, we make use of the Taylor expansions for $\bar{m}$ and $\bar{p}$:

\begin{equation}\label{tayM}
    \bar{m}(\bar{r}) = \bar{\rho}(\bar{p}_0)\bar{r}^3 + O(\bar{r}^5),
\end{equation}
\begin{equation}\label{tayP}
    \bar{p}(\bar{r}) = \bar{p}_0 + \frac{\bar{p}_2}{2} \bar{r}^2 + O(\bar{r}^3),
\end{equation}
with
\begin{equation}
    \bar{p}_2 = -G_\star\left[\bar{\rho}(\bar{p}_0)+3\bar{p}_0\right]\left[\bar{\rho}(\bar{p}_0)+\bar{p}_0\right].
\end{equation}
Then, given an EoS of the form $\bar{\rho}=\bar{\rho}(\bar{p})$, the family of solutions is parametrized by the central pressure $\bar{p}_0$, which is equivalent to get it parametrized by the central density $\bar{\rho}(\bar{r} = 0)=\bar{\rho}_0$.
To close the TOV system an EoS is needed, to describe the matter content of the spherical halo.

\subsection{Velocity profile and phenomenological EoS for dark matter}

\begin{figure}
    \centering
    \includegraphics[width=0.49\textwidth]{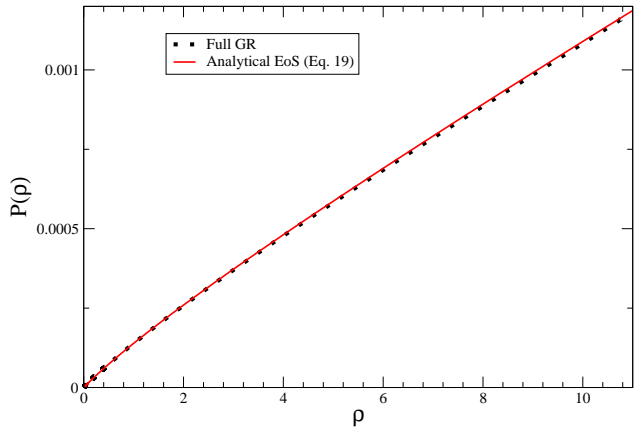}
    \caption{Comparison of the EoS obtained using the full general relativistic approach (dotted line) against the approximate solution obtained in the Newtonian approximation given by equation \ref{EOS_P}.}
    \label{fig1:EOS}
\end{figure}

The quest to determine the nature of dark matter is perhaps one of the most challenging problems in modern physics. The particle physics approach is the most dominant in the literature. It consists of proposing a dark matter candidate that heals some standard particle model problems as well as providing a viable weakly interacting particle that plays the role of dark matter. Well-motivated examples of such dark matter candidates are the neutralino (the lightest stable neutral supersymmetric particle) or the axion (a pseudoscalar boson that solves the Strong CP problem).   

On the other hand, an approach where stellar dynamics and a collection of reduced hypotheses determine the general properties of dark matter is not mainstream. Fluid dark matter models represent an example of such a phenomenological approach. 
In particular, the rotational velocity profiles of galaxies can provide important insights about dark matter under this fluid approach. 
For instance, in \cite{Barranco:2013wy} several velocity rotational profiles of galaxies were considered to construct EoS for dark matter. 
Actually, what is shown in \cite{Barranco:2013wy} is that if dark matter behaves as a perfect fluid, imposing the spacetime to be spherically symmetric and static, and given a profile of rotational velocities $v_t(r)$ of stars in a galaxy, thus,  dark matter halos in hydrostatic equilibrium demand an effective EoS. The argument is simple enough: for a spherically symmetric and static spacetime, for test particles in circular motion, there is a relationship between tangential velocity and the gravitational potential $\Phi$, given by $\Phi'(r)=v_t^2(r)/r$ (see for instance \cite{Rahaman:2010xs,Nunez:2010ug,Gong:2020lev}). Since the gravitational potential is fixed once $v_t(r)$ is known, equations (\ref{TOVmass}) and (\ref{TOVpressure}) can be combined to give a first-order differential equation for $p$ as a function of $\rho$. Thus, for every phenomenological velocity $v_t(r)$ a phenomenological EoS, $p=p(\rho)$, can always be derived.

A particular profile used in \cite{Barranco:2013wy} was the one presented by Persic, Salucci, and Stel \citep{Persic1996}. This rotational velocity profile called the PSS profile (Persic, Salucci, Stel) or the Universal Velocity profile, has the analytical expression:
\begin{equation}
    \frac{v_t^2(\bar r)}{c^2}=\beta^2(\bar r) = \beta_0^2~ \frac{\bar{r}^2}{(\bar{r}^2+a^2)},\label{PSSprofile}
\end{equation}
where $\beta_0$ and $a$ are parameters that need to be observationally determined for each galaxy. 
As explained above, given the velocity profile equation (\ref{PSSprofile}), the gravitational potential $\Phi$ can be computed and then it is possible to obtain the corresponding profiles for mass, density and pressure \citep{Barranco:2013wy}, which in the Newtonian regime have the analytical expressions\footnote{Considering the size of the halos and if dark matter is a particle at galactic scales, then the Newtonian regime is an excellent approximation.}:
\begin{equation}\label{PSSm}
    \bar{m} = \frac{\beta_0^2}{G_\star}\frac{\bar{r}^3}{(\bar{r}^2+a^2)},
\end{equation}
\begin{equation}\label{PSSrho}
    \bar{\rho} = \frac{\beta_0^2}{3G_\star}\frac{(\bar{r}^2+3a^2)}{(\bar{r}^2+a^2)^2},
\end{equation}
\begin{equation}\label{PSSp}
    \bar{p} = \frac{\beta_0^4}{6G_\star}\frac{(\bar{r}^2+2a^2)}{(\bar{r}^2+a^2)^2}.
\end{equation}
These represent the mass, density and pressure profiles for the dark matter halo that induces a rotational velocity profile given by equation (\ref{PSSprofile}). Observe that given the radial profiles  (\ref{PSSrho}) and (\ref{PSSp}), for $\bar r=0$ we have 
\begin{equation}\label{EOSrho}
    \bar{\rho}(\bar r=0)=\bar{\rho}_\bullet = \frac{\beta_0^2}{G_\star a^2},\quad \bar{p}(\bar r=0)=\bar{p}_\bullet = \frac{\beta_0^4}{3G_\star a^2},
\end{equation}
and we can combine equation (\ref{PSSrho}) with  (\ref{PSSp}) to obtain the corresponding EoS for dark matter with the PSS velocity profile:
\begin{equation}
    \bar{p}(\bar{\rho}) = \bar{p}_\bullet\left[\frac{3}{4}\frac{\bar{\rho}}{\bar{\rho}_\bullet}-\frac{1}{16}\left(1-\sqrt{1+24\frac{\bar{\rho}}{\bar{\rho}_\bullet}}\right)\right], \label{EOS_P}
\end{equation}
where $\bar{\rho}_\bullet$ and $\bar{p}_\bullet$ are related to $\beta_0$ and $a$ through equations \ref{EOSrho}.\\

This EoS \ref{EOS_P} has a barotropic limit for $\bar{\rho}\ll\bar{\rho}_\bullet$ given by $\bar{p}=\frac{3\bar{p}_\bullet}{2\bar{\rho}_\bullet}\bar{\rho}$.\footnote{There is a typo in this formula in \cite{Barranco:2013wy}.} For later use, we need to invert the relation \ref{EOS_P} to have the density as a function of the pressure, such that gives us the EoS studied in this work, equation \ref{EOS_rho}.

As we have mentioned, the analytical expression for the EoS obtained by the PSS velocity profile was derived in the Newtonian approximation. Nevertheless, the procedure to compute the EoS can be performed within the framework of General Relativity. In order to do that we refer to the appendix in \cite{Barranco:2013wy}. Unfortunately, in this case, there is no analytical expression for the mass, density, and pressure profiles. Thus, the resulting EoS, in the full general relativistic approach, can be computed only by numerical methods.
Before proceeding to build self-gravitating configurations solving the TOV system with this particular EoS for the dark matter, it is important to check if there are important differences in the EoS obtained numerically in a General Relativity treatment for the PSS rotational velocity profile and the EoS given by equation (\ref{EOS_rho}).
This comparison is shown in Figure \ref{fig1:EOS}. We can see that there are no significant differences and thus for the rest of our work we use equation (\ref{EOS_rho}).

\section{Properties and characteristics of dark matter halos with EoS from the universal velocity profile}\label{section_3}

In this section, we solve the TOV system with the EoS given by equation \ref{EOS_rho}, in order to find the self-gravitating dark matter halos. We follow \cite{Barranco:2013wy} in the hypothesis that dark matter does not interact with the baryonic matter, and that the latter acts only as tracers of the gravitational potential created by the dark matter in the halo. Therefore, the only gravitating structure that needs to be determined is the dark matter halo.
There are two free parameters in the dark matter EoS equation \ref{EOS_rho}: $\bar \rho_\bullet$ and $\bar p_\bullet$. Such EoS was indeed derived with  $\bar{\rho}(\bar{r}=0)=\bar{\rho}_\bullet$ and $\bar{p}(\bar{r}=0)=\bar{p}_\bullet$, but this is a very particular choice of initial conditions for the TOV system. In general, $\bar{\rho}(\bar{r}=0)\ne\bar{\rho}_\bullet$ and $\bar{p}(\bar{r}=0)\ne \bar{p}_\bullet$, and this is the reason why we use different symbols for $\bar{\rho}_\bullet$ and $\bar{\rho}_0$ and $\bar{p}_\bullet$ and $\bar{p}_0$. 
We consider $\bar{\rho}_\bullet$ and $\bar{p}_\bullet$ as fixed quantities and construct the family of halos with varying $\bar{p}_0$. Once ($\bar{\rho}_\bullet$,$\bar{p}_\bullet$) are fixed, the TOV system is closed and it is possible to find all possible self-gravitating configurations. As free data, we can either choose $\bar p_0$ or its equivalent $\bar \rho_0$, since they are related through equation \ref{EOS_rho}.  
Before continuing, we must choose the values ($\bar{\rho}_\bullet$,$\bar{p}_\bullet$) that we explore in this work. In \cite{Barranco:2013wy}, 
20 galaxies were fitted, and each galaxy demanded a different value for ($\bar{\rho}_\bullet$,$\bar{p}_\bullet$). 
We can consider some of those values as a starting point. We concentrate on the values obtained for the three galaxies presented in Table \ref{tablaParametros}, as the results that we present later are qualitatively similar and cover the range of $\bar{\rho}_\bullet$ and $\bar{p}_\bullet$ that fits a representative number of galaxies studied in \cite{Barranco:2013wy}.

\begin{table}
    \caption{Parameters of the selected galaxies}
    \centering
    \begin{tabular}{c c c c c c}
        \hline
        Label  & Galaxy & $\beta_0 (10^4)$ & $a$  & $\bar{\rho}_\bullet$   & $\bar{p}_\bullet$ \\\hline
        A & U5750      & 3.23             & 7.75 & $3.63\times10^{-3}$    & $1.26\times10^{-10}$ \\
        B & ESO2060140 & 4.00             & 2.16 & $7.17\times10^{-2}$    & $3.82\times10^{-9}$ \\
        C & U11748     & 7.94             & 1.07 & $1.15$                 & $2.42\times10^{-7}$ \\\hline
    \end{tabular}
    \label{tablaParametros}
\end{table}

\begin{figure}
    \centering
    \includegraphics[width=0.48\textwidth]{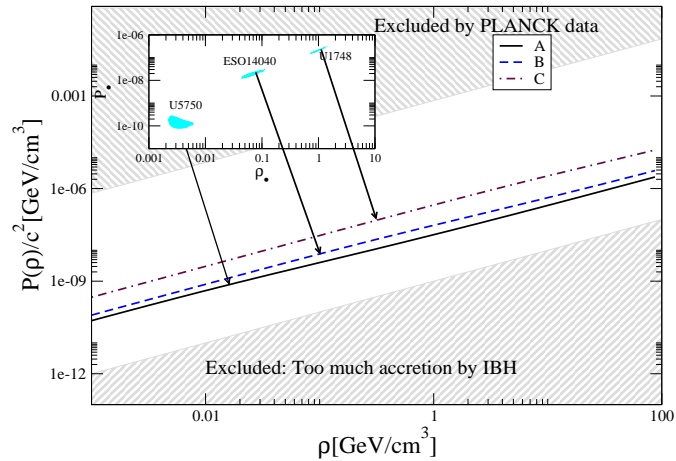}
    \caption{By fixing ($\bar{\rho}_\bullet, \bar{p}_\bullet$) with the central values of the fit in equation (\ref{EOS_rho})  three different EoS $\bar P(\bar \rho)$
    are obtained. Those EoS are plotted and compared its behavior with current exclusion areas obtained by analyzing the CMB anisotropies \cite{Xu:2013mqe} and dark matter accretion by Intermediate Mass Black Holes \cite{Pepe:2011dk,Lora-Clavijo:2014kha}. Inset plot: Contours at 90\% C.L. in the ($\bar{\rho}_\bullet, \bar{p}_\bullet$) plane that fits the rotational curve velocities of the galaxies listed in Table \ref{tablaParametros}.}
    \label{fig:status}
\end{figure}

In Fig. \ref{fig:status}, the resulting EoS for each pair  ($\bar{\rho}_\bullet$,$\bar{p}_\bullet$) are plotted. The inset plot shows the allowed region in the space ($\bar{\rho}_\bullet$,$\bar{p}_\bullet$) that fits the rotational velocity data of the mentioned galaxies in Table \ref{tablaParametros} at 90\% of C.L. These three galaxies cover most of the relevant region that fits most of the 20 galaxies studied in  \cite{Barranco:2013wy}. 
The resulting dark matter EoS that we explore are within the allowed region that is not excluded either by analysis done using cosmological data \citep{Xu:2013mqe} or by studies of accretion of dark matter by Intermediate Mass Black Holes \citep{Pepe:2011dk,Lora-Clavijo:2014kha}.

\subsection{EOS with fixed parameters}

In the present work, we interpret equation \ref{EOS_rho} in two ways. First, in this section, we consider that $\bar{\rho}_\bullet$ and $\bar{p}_\bullet$ are constants valid for all possible dark matter halos, therefore there is a unique EoS for a dark matter where the only relevant variables are the density and pressure. Once we have such an EoS, we integrate the TOV equations, varying the central pressure $\bar p_0$, obtaining a family of dark matter objects. 

\begin{figure*}[!b]
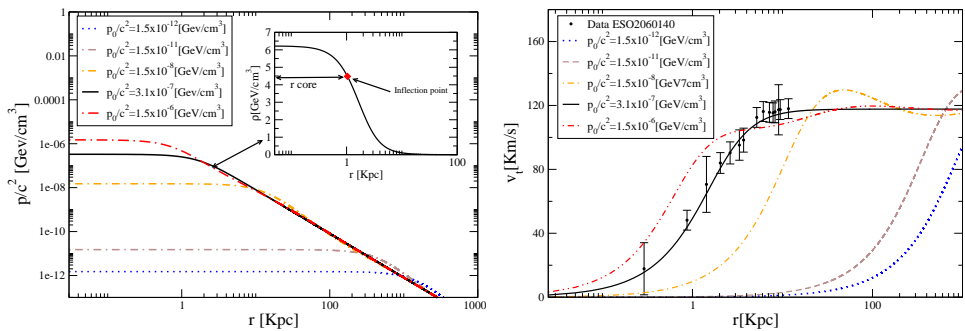

    \centering
    \includegraphics[width=0.45\textwidth]{perfiles_plot.eps}
    \includegraphics[width=0.43\textwidth]{one_family_nofit.eps}
    \caption{Left panel: Radial pressure profiles obtained as solutions of the TOV equations with dark matter EoS labeled as EoS B in Table \ref{tablaParametros}. Each curve corresponds to a different initial value for $\bar p_0$. The inset plot shows the equivalent radial density profile $\rho(r)$ obtained by mapping the pressure profiles via equation \ref{EOS_rho} for the particular case where $\bar p_0=\bar p_\bullet$. 
    The red point corresponds to the point where the density profile has an inflection point and we define the core radius of the configuration as the radius where this inflection point occurs.  Right panel:  The rotational velocity profile 
    for different halos (different initial values of $p_0)$ within one family of solutions. It can be observed that from all those configurations only the configuration where $\bar p_0=\bar p_\bullet$ can fit the observed rotational velocities for the galaxy ESO2060140.}
    \label{velT}
\end{figure*}

With compact objects obtained from the EoS through the TOV equations, an important feature is the M-R (mass-radius) diagram. This diagram gives an idea of the typical sizes of possible objects and a criterion for the stability of the configurations. For barotropic EoS, and also for the EoS \ref{EOS_P}, there is the problem of defining the radius of the object, since the pressure and density never become zero, and the mass is divergent. Specifically, for a barotropic EoS, the pressure decreases quadratically with $r$, with the density decreasing also quadratically, and therefore the mass increasing linearly. 

As the pressure never becomes zero there is not a clear way of defining the size of the object.

As we have mentioned earlier, for $\bar p \ll 1$ the EoS given by equation \ref{EOS_P} has a barotropic limit, which is consistent with the feature of flat rotation curve profile of the galaxies for large values of $\bar r$ and can not be avoided in the present setting. To overcome this problem, we consider the boundary of the halo to be located where the density of dark matter becomes the density of dark matter in between galaxies. That is, we consider the radius at which it is no longer possible to distinguish between the halo and the dark matter background. For this, we take the value of \citep{Planck:2018vyg}:
\begin{equation}
    \rho_m = 1.2\times 10^{-6}\,GeV/cm^3,
\end{equation}
for the intergalactic dark matter density, which in our dimensionless variables reads as
\begin{equation}
    \bar{\rho}_m = 1.3\times 10^{-8}.
\end{equation}
Therefore, we consider the radius $R$ and mass $M$ of the dark matter object as the radius and mass where the dark matter density is equal to $\rho_m$.

The velocity profile given by equation \ref{PSSprofile} has been proposed because it has constant rotational velocities for $\bar r \gg 1$ and cored galaxies. Thus, it is natural to expect that the resulting density profile from the solution of the TOV system with EoS given by equation \ref{EOS_rho} has a core.
We consider the "core" of the object as the radius where the density profile has an inflection point and denote such radius as $R_{core}$ and the corresponding mass as $M_{core}$.\\

\begin{table*}
    \caption{Maximum mass for the dark matter halos and related parameters.}
    \centering
    \begin{tabular}{c c c c c c c}
        \hline
        EoS & $M\,[10^{10}M_\odot]$ & $M_{core}\,[10^{10}M_\odot]$ & $R\,[kpc]$ & $R_{core}\,[kpc]$ & $\bar{\rho}_0$ & $\bar{p}_0$ \\\hline
         A & $756$ & $45.1$ & $2851$ & $668$ & $1.86\times 10^{-7}$ & $9.7\times 10^{-15}$ \\
         B & $1437$ & $86$ & $3531$ & $828$ & $1.86
         \times 10^{-7}$ & $1.49\times 10^{-14}$ \\
         C & $11242$ & $672$ & $7009$ & $1644$ & $1.86\times 10^{-7}$ & $5.9\times 10^{-14}$ \\\hline
    \end{tabular}
    \label{tablaCritica}
\end{table*}
\begin{figure*}
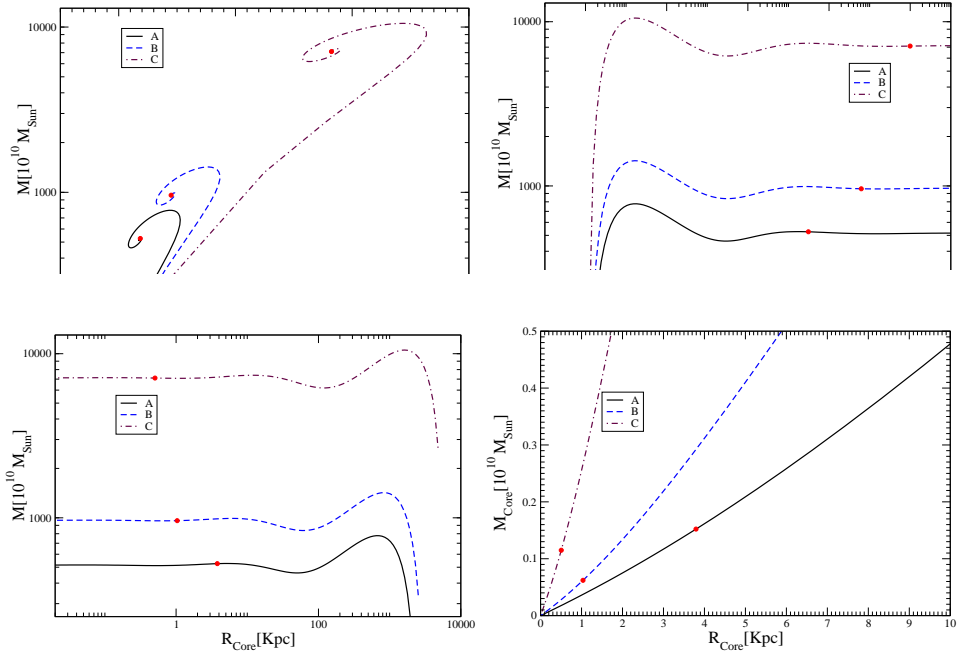

\gridline{\fig{MT_RT.eps}{0.45\textwidth}{(a)}
          \fig{MT_Rho.eps}{0.45\textwidth}{(b)}}
\gridline{\fig{MT_RC.eps}{0.45\textwidth}{(a)}
          \fig{MC_RC.eps}{0.45\textwidth}{(b)}}          
\caption{Main properties of the dark matter halos obtained with the A, B, C EoS. From left to right and from up to down: 
    The total mass $M$ as a function of the radius of the halo $R$. The total mass $M$ as a function of the central density $\rho(0)$. 
    The total mass $M$ as a function of the core radius $R_{core}$ and finally, the mass of the core as a function of the core radius. With a red point the configurations that fit the observed rotational curves of the galaxies enumerated in Table \ref{tablaParametros} are shown. In all cases, the red points are in the unstable branch.}
\label{fig:MTRT}
\end{figure*}

To perform the numerical integration of equations \ref{TOV1}-\ref{TOV2} we use the Runge-Kutta-Felhberg (4,5)  algorithm already implemented in SageMath \citep{SageMath}. Due to the system of ODEs being formally singular at $\bar{r}=0$, we use the Taylor expansions \ref{tayM}-\ref{tayP} to transport the initial conditions to the first point of the integration grid. We use a non-uniform grid, due to the relatively fast change of variables close to $\bar{r}=0$ concerning the regions far away from the center. The $k$$-th$ point of the integration grid is
\begin{equation}
    \bar{r}_k = \bar{r}_n \left(\frac{k}{n}\right)^{1.2},\quad k = 1,\ldots,n,
\end{equation}

\noindent
being $n$ the number of points on the grid, and $\bar{r}_n$ the upper limit for the integration. We use $n = 10^5$.
Examples of the solutions can be seen in Figure \ref{velT}. The left panel shows radial pressure profiles obtained as solutions of the TOV equations with the dark matter EoS labeled as EoS B in Table \ref{tablaParametros}. Each curve corresponds to a different initial value for $\bar p_0$.
A black solid line corresponds to the solution with $\bar p_0=\bar p_\bullet$. 
The inset plot shows the equivalent radial density profile $\rho(r)$ obtained by mapping the pressure profile via equation \ref{EOS_rho} for this particular case where $\bar p_0=\bar p_\bullet$. The red diamond corresponds to the point where the density profile has an inflection point and we define the core radius of the configuration as the radius where this inflection point occurs.  
Besides the radial pressure profile $\bar p(\bar r)$, the mass profile $\bar m(\bar r)$ is obtained as well as the solution of the TOV system. 
Thus the rotational velocity profile can be directly computed by $v_t=\sqrt{\frac{G m(r)}{r}}\,.\label{vt}$ 

The right panel of Fig. \ref{velT} shows the rotational velocity profiles for the different halos that correspond to the
pressure profiles from the left panel.  
It should be noted that although the EoS was obtained from the PSS velocity profile, in the family of solutions, the only object that fits this profile is the one that has exactly the values $\bar{\rho}_0=\bar{\rho}_\bullet$ and $\bar{p}_0=\bar{p}_\bullet$. To make this point explicit, in Figure \ref{velT} we have plotted the velocity profiles of five halos in the same family. We have used a logarithmic scale on the axis of abscissas to make the differences in the profiles more obvious, remembering that the only PSS profile is the one with $\bar{p}_0 = 3.8\times 10^{-9}$.\\

In Figure \ref{fig:MTRT}, the diagrams for mass vs radius, mass vs central density, mass vs core radius and core mass vs core radius, respectively, are presented, using the parameters ($\bar{\rho}_\bullet, \bar{p}_\bullet$) shown in Table \ref{tablaParametros} for the EoS obtained from the galaxies U5750, ESO2060140 and U11748, now labeled as EoS A, B and C. We see that for the three pairs of $(\bar{\rho}_\bullet,\bar{p}_\bullet)$ here considered, the results are qualitatively the same. The mass has a maximum, indicating the existence of a stable branch and an unstable branch. Here we use the criteria that a static object of perfect fluid can only pass from stability to instability concerning some particular radial normal mode at a value of the central density where the mass is an extreme \citep{Weinberg1972}. The maximum mass, and the corresponding radius, core radius, and core mass are summarized in Table \ref{tablaCritica}. It is of interest to note that the value of $\bar{\rho}_0$ for the maximum mass coincides with the third digit for the three families of galaxies. It seems to indicate that the deciding factor for stability is the central density; objects with lower central densities are stable, while objects with higher central densities are unstable. In Figure \ref{fig:MTRT}, the halos obtained with the value $\bar{p}_0=\bar{p}_\bullet$, which are the only ones that fit the corresponding observed rotational velocities of the galaxies, are depicted with a red dot, and in all cases, they are definitely unstable. Similarly, in Figure \ref{velT} the profile for $\bar{\rho}_0 \le 1.86\times 10^{-7}$ correspond to a stable halo object while the others are unstable. 

\begin{figure}
    \centering
    \includegraphics[width=0.49\textwidth]{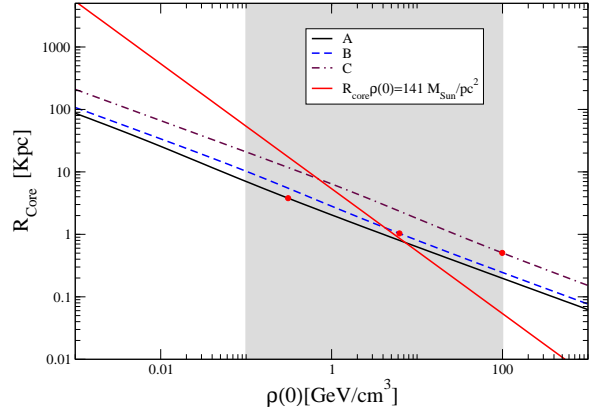}
    \caption{$R_{core}$ as a function of $\rho(0)$ for the EoS with fixed parameters ($\bar p_\bullet,\bar \rho_\bullet$) studied in this work. 
    The values of the core radius of the halos that fit the observed rotational velocities are of the order of a few kiloparsecs, while the core radius of the stable configurations, with $\bar \rho_0< 1.86 \times 10^{-7}$, are of the order of a thousand kiloparsecs. The red solid line corresponds to the relation $R_{core}\rho_0=141 M_\odot/$pc$^2$ found in \cite{Donato:2009ab,Gentile:2009bw}.}
    \label{fig:Rcore}
\end{figure}

Figure \ref{fig:MTRT} summarizes the global general properties of dark matter halos that can be obtained with the EoS derived from the PSS velocity profile. The core radius is an important one. Note that the core radius for stable configurations is bigger than 100 Kpc. Thus, the stable configurations are unable to fit the observed rotational curves, as they demand cores of the order of a few kiloparsecs. This issue can be seen graphically on the right panel of Figure \ref{velT}. 
Following the analysis of the core radius of the resulting self-gravitating objects that model our dark matter halos, in Figure \ref{fig:Rcore} we have plotted the core radius as a function of the central density. Here again, the red dots indicate the values that best fit the rotational curves of galaxies shown in Table \ref{tablaParametros}. 
Furthermore, the solid red line shown in Figure \ref{fig:Rcore} corresponds to the observational evidence that the central surface density, defined as the product of the central density times the core radius, of galaxy dark matter halos, is nearly constant and independent of galaxy luminosity \citep{Donato:2009ab,Gentile:2009bw}. 
Observe that in general,  A, B and C EoS studied in this work do not follow this universal relation of constant surface density. It is worth noting that the best-fit point for the galaxy ESO2060140 lies near the red solid line, this motivates us to explore another possible set of EoS as equation \ref{EOS_rho}.

\subsection{Constant ($\rho_0 R_{core}$)}
\begin{figure}
    \centering
    \includegraphics[width=0.49\textwidth]{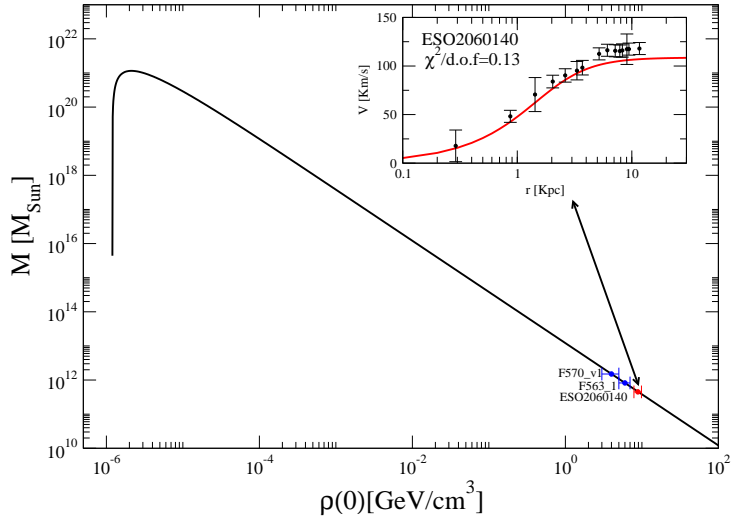}
    \caption{Plot of $\bar m_m(\bar \rho_0)$ with constant $\bar p_\bullet$ and variable $\bar \rho_0=\bar \rho_\bullet$. The maximum mass is reached for $\bar\rho_{0,crit}=2.7\times 10^{-8}$; all configurations to the right are unstable. The value of $\bar \rho_0$ needed to fit the rotational curve of ESO2060140 is bigger than this value, thus the configuration that fits the data is in the unstable branch.  }
    \label{trouble}
\end{figure}

\begin{figure*}
\gridline{\fig{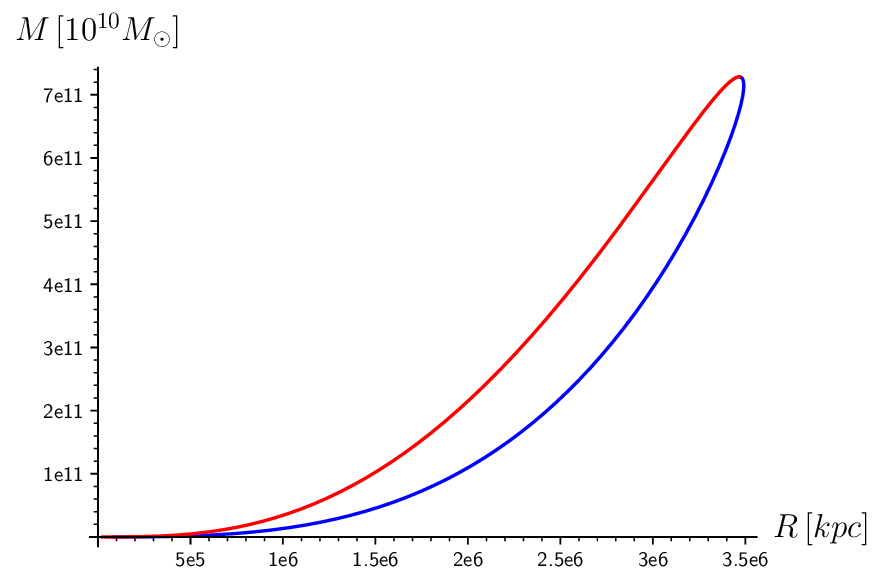}{0.3\textwidth}{(a)}
          \fig{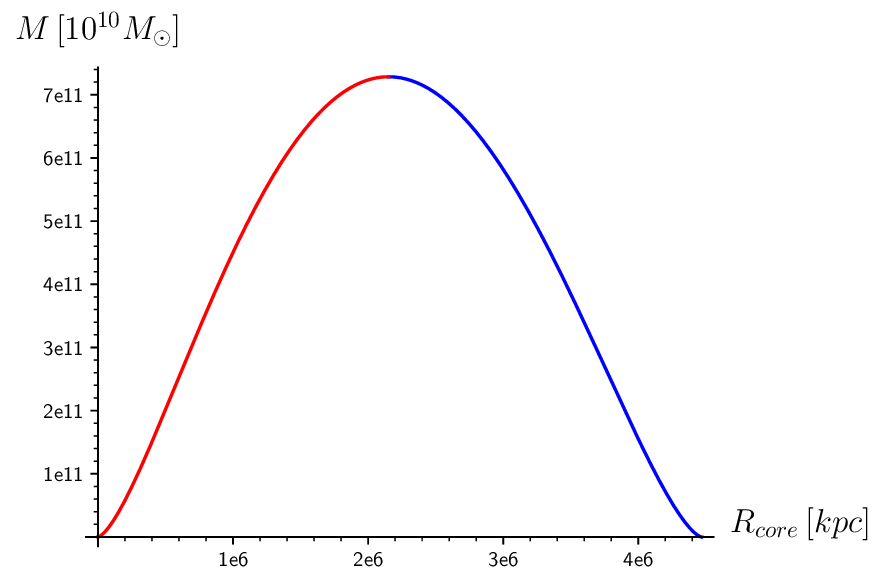}{0.3\textwidth}{(b)}
          \fig{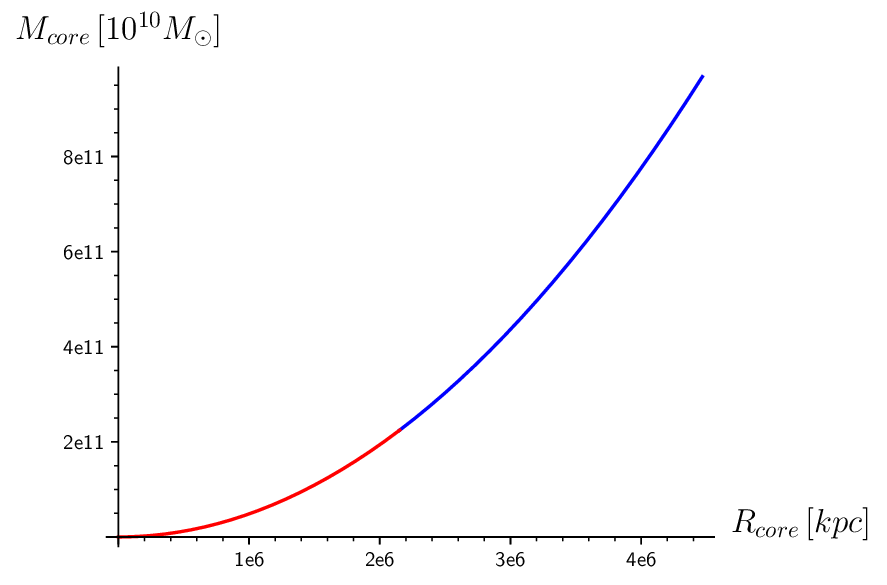}{0.3\textwidth}{(c)}}
\caption{$M-R$, $M-R_{core}$ and $M_{core}-R_{core}$ diagrams for the EoS with constant $\bar{p}_0$.}
\label{figP0cte}
\end{figure*}

Our previous results reveal a fundamental problem when the EoS given by equation \ref{EOS_rho} is used to model dark matter halos: the halo that fits the rotational curve belongs to the unstable branch of possible configurations. 
This could be an effect of the particular (and arbitrary) selection of the sets of parameters ($\bar \rho_\bullet,\bar p_\bullet$) that we have done. 
In this section, we interpret the EoS differently. We assume that all galaxies follow the PSS rotational velocity profile, which gives rise to the dark matter EoS \ref{EOS_rho}, but they do not have the same $\bar{\rho}_\bullet$. Instead of that, following the work \cite{Donato:2009ab,Gentile:2009bw}, we consider that the product $\rho_0 ~R_{core}$ has the constant value
\begin{equation}\label{valUniv}
    \rho_0~R_{core} = 141^{+82}_{-52}\,M_\odot/pc^2.
\end{equation}
The core radius is where the density profile has an inflection point and from \ref{PSSrho},
\begin{equation}
    \bar{r}_{core} =\sqrt{\frac{2\sqrt{34}-11}{3}}a= \sqrt{2\sqrt{34}-11}\sqrt{\frac{\bar{p}_\bullet}{G_\star}}\frac{1}{\bar{\rho}_0},
\end{equation}
which together with \ref{valUniv} implies
\begin{equation}\label{valP0}
    \bar{p}_0 = \bar p_\bullet=\frac{G_\star(\bar{\rho}_0~\bar{r}_{core})^2}{2\sqrt{34}-11} = 2.5\times 10^{-9}.
\end{equation}

Now, we interpret \ref{EOS_rho} in the following way. The parameter $\bar{p}_\bullet$ in \ref{EOS_rho} is a constant for all galaxies, having the value \ref{valP0}. The parameter $\bar{\rho}_\bullet$ is both the corresponding parameter in \ref{EOS_rho} and the central density of the galaxies $\bar{\rho}_0$, and it is not a constant for different galaxies. Therefore, we have a family of dark matter objects, parameterized by $\bar{\rho}_0$, each galaxy with an EoS with its own set of parameters, being $\bar{p}_\bullet$ a constant. Although having a different EoS for each galaxy may seem like the model loses its value, the interpretation is not that each galaxy possesses a different type of dark matter. Instead of that, we consider that there is another parameter, the simplest one being the temperature, which has not been considered here. In this sense, having EoSs with different parameters amounts to having galaxies with different temperature profiles. The interpretation is the same as with the classical Eddington model for stars made of an ideal gas, where the EoS is that of an ideal gas, but the temperature profile is supposed to be such that the pressure contribution from the gas has a constant ratio concerning the radiation pressure. In this way, the equations for $p$ and $m$ can be integrated without explicitly considering the temperature, and each star has a particular EoS in the form $\rho(p)$.

With this interpretation in mind, there is only one family of solutions with only one free parameter: the central density $\bar \rho_0$. The values of $R$, $R_{core}$, $M$ and $M_{core}$ have the same meaning as before, but now they can be obtained analytically as functions of central density $\bar{\rho}_0$, from \ref{PSSm} and \ref{PSSrho}. Then, in terms of the dimensionless quantities, we have the mass of the core:
\begin{equation}
    \bar{m}_{core} =\bar m(\bar r_{core})= \frac{3(2\sqrt{34}-11)^\frac{3}{2}}{2(\sqrt{34}-4)}\frac{\bar{p}_\bullet^\frac{3}{2}}{G_\star^\frac{3}{2}\bar{\rho}_0^2}.
\end{equation}
The dark matter halo ends at the point where $\bar \rho$ given in equation \ref{PSSrho} is equal to $\bar \rho_m$, and therefore the radius of the configuration is
\begin{equation}
    \bar{r}_m = \sqrt{\frac{\bar{p}_{\bullet}}{2 \, G_\star \bar{\rho}_{m}\bar{\rho}_{0}}}   \sqrt{1 - 6 \, \frac{\bar{\rho}_{m}}{\bar{\rho}_0} + \sqrt{1 + 24 \, \frac{\bar{\rho}_{m}}{\bar{\rho}_0}}},
\end{equation}
and the total mass of the halo $\bar m_m$ is obtained with $\bar m(\bar r=\bar r_m)$, i.e.
\begin{equation}
    \bar{m}_m = \frac{3 \bar{p}_{\bullet}^{\frac{3}{2}}}{\sqrt{2}G_\star^{\frac{3}{2}}\sqrt{\bar{\rho}_{m}}\bar{\rho}_{0}^{\frac{3}{2}}} \frac{ {\left( 1 - 6 \,  \frac{\bar{\rho}_{m}}{\bar{\rho}_0} + \sqrt{1 + 24 \, \frac{\bar{\rho}_{m}}{\bar{\rho}_0}}  \right)^\frac{3}{2}} }{1 + \sqrt{1 + 24 \, \frac{\bar{\rho}_{m}}{\bar{\rho}_0}} }. \label{mass_salucci}
\end{equation}
Equation \ref{mass_salucci} gives the mass of the halo as a function of the central density. We have plotted $\bar m(\rho_0)$ in Figure \ref{trouble}. We see that the mass has a maximum value, which from equation \ref{mass_salucci}  
corresponds to the central density
\begin{equation}
    \bar{\rho}_{0,crit} =(3\sqrt{19}-11)\bar\rho_m= 2.7\times 10^{-8},
\end{equation}
being the maximum mass
\begin{equation}
    M_{crit} = \frac{3\sqrt{951029+220742\sqrt{19}}\bar p_\bullet}{12500 G_\star^{3/2}\bar \rho_m^2}=7.3\times 10^{21}M_\odot,
\end{equation}
and the corresponding object radius
\begin{equation}
    R_{crit} = \frac{232+61\sqrt{19}}{625\bar \rho_m}\sqrt{\frac{\bar p_\bullet}{2 G_\star}}=3.5\times 10^6 kpc.
\end{equation}

Although it is possible to obtain analytical expressions for these quantities in terms of one of the others, for instance, $\bar{m}_m(\bar{r}_m)$, the formulas are not very enlightening and we prefer to consider them as given parametrically through $\bar{\rho}_0$. For the configuration with maximum mass, the core radius is
\begin{equation}
    R_{core,crit} = 2.2\times 10^6 kpc,
\end{equation}
being the core mass
\begin{equation}
    M_{core,crit} = 2.2\times 10^{21}M_\odot.
\end{equation}

In Figure \ref{figP0cte}, the relationship between mass, core mass, radius, and core radius, is shown. 
The maximum mass represents, as before, the configuration that divides stable from unstable configurations. 

The stable branches are those objects with 
\begin{equation}
    \bar{\rho}_0 <\bar{\rho}_{0,crit},\quad M<M_{crit},\
\end{equation}
\begin{equation}
    M_{core}>M_{core,crit},\quad R_{core}>R_{core,crit},
\end{equation}
while the unstable branch is the one with
\begin{equation}
     \bar{\rho}_0 >\bar{\rho}_{0,crit},\quad M<M_{crit},
\end{equation}
\begin{equation}
    M_{core}<M_{core,crit},\quad R_{core}<R_{core,crit}.
\end{equation}
In Figure \ref{figP0cte}, the unstable branch is indicated by the red curve, while the stable branch by the blue one.

As expected, we can fit some rotational velocity profiles with this set of configurations. 
The fit can be done through a $\chi^2$ analysis of the rotational curves measured in some Low Surface Brightness Galaxies \citep{deBlok:2002vgq} and the theoretical curve obtained by equation \ref{vt}, with the mass computed by equation \ref{mass_salucci}. There is only one free parameter, $\bar \rho_0$, and thus we minimize $\chi^2(\bar \rho_0)=\sum_i \frac{(v_t^i-v_t(\bar \rho_0))^2}{(\delta v_t^i)^2}$ in order to find the best fit point. Here $v_t^i$ are the observed data and $\delta v_t^i$ are the errors in the data points reported in \cite{deBlok:2002vgq}. 
The particular case of the rotational velocity data from ESO2060140 is shown as an inset plot in Figure \ref{trouble}. It can be seen 
that the fit is reasonably good ($\chi^2_{min}/\text{d.o.f.}=0.13$). The best-fit value for this galaxy is $\bar \rho_0=8.87$ GeV/cm$^3$,  and it is plotted in Figure \ref{trouble} as a red point. Thus, this dark matter halo lies on the unstable branch.  

Other galaxies also can be fitted by fixing $\bar p_\bullet$, as has been explained above. For instance, the galaxies F563-1 and F570 v1,
and the best-fit points for those galaxies are plotted as blue points in Figure \ref{trouble}.
One more time, those configurations are in the unstable branch.

\begin{table*}
    \centering
    \caption{Gravitational energy and related parameters for the last stable configurations.}
    \begin{tabular}{c c c c c c c}
        \hline
        Galaxy & $M\,[10^{10}M_\odot]$ &  $R\,[\mbox{kpc}]$  & Gravitational Energy \,$[Mc^2]$ & Density\,$[gr/cm^3]$&  Surface gravity \\
        \hline
         U5750 & $756$ &  $2851$ &  $1.27\times 10^{-7}$ & $5.27\times 10^{-30}$ & $1.29 \times 10^{-13}$ \\
         ESO2060140 & $1437$  & $3531$ &  $0.002$
          & $5.27\times 10^{-30}$ & $1.61 \times 10^{-13}$ \\
          U11748 & $11242$  & $7009$ & $0.080$ & $5.99\times 10^{-27}$& $3.47 \times 10^{-11}$ \\
          \hline
    \end{tabular}
    \label{tablaEnergy}
\end{table*}

\section{Discussion}\label{section_4}

We have integrated the TOV system with the dark matter EoS given by equation \ref{EOS_rho}. Four cases were analyzed: three families of solutions were obtained for the representative sets of parameters ($\bar p_\bullet, \bar \rho_\bullet$) given in Table \ref{tablaParametros}, as they cover most of the parameter space that can fit the observed rotational velocities of Low Surface Brightness Galaxies. The fourth family of solutions was obtained by fixing $\bar p_\bullet$ for the constancy of the surface brightness in galaxies found in \cite{Donato:2009ab,Gentile:2009bw} to be always satisfied. This last family of solutions has as a free parameter the central density.
The four families of dark matter halos have similar properties and one configuration in all cases is of special interest: the configuration with maximum mass. The parameters presented in table \ref{tablaCritica} represent the maximum mass and corresponding radii for our galaxy halos. The stable configurations are to the right of the maximum and the unstable ones to the left in the $M-R$ diagrams. As is common in various mass-radius relations, for large values of the central pressure the curves start to curl. The maximum in the curves divides the stable configurations from the unstable ones. The halos that fit the rotational curve for the considered galaxies are all on the unstable branch. It is important to note that this separation into stable/unstable objects can not be forced \textit{a priori} into the EOS. With the assumption made in \cite{Barranco:2013wy} the EOS is unique, and the consistency check at that stage not to discard the EOS was that $p$ is a monotonically increasing function of $\rho$. Nonetheless, it is possible to engineer an EOS such that objects with the characteristics of DM halos are stable, for example with a polytropic EOS, but then the velocity profile does not coincide with the PSS profile.

In table \ref{tablaEnergy} we have estimations for the gravitational energy, the density, and the surface gravity of the halos of our set of three galaxies. We found that, due to the gravitational energy being not an appreciable fraction of the rest energy, the halos are not considered relativistic objects. Even more, the mean density is low, in the order of the current density of the Universe of $9.9 \times 10^{-30}$ gr/cm$^3$, in correspondence to a flat universe.  However, U11748 exhibits a higher density, about $6\times 10^{-27}$ gr/cm$^3$, so it can expand openly. This happens because the stable configurations need to have a central density which is only a few times the average density of the universe. Thus the dark matter halos have a radius of the order of thousands of kiloparsecs, the core radius of hundreds of kiloparsecs, and therefore stable halos obtained through equation \ref{EOS_rho} are not good models for realistic halos that fit the rotational curve of galaxies. It is known that CDM is capable of producing reasonable fittings to DM density profiles, in this sense, the stable configurations obtained from equation \ref{EOS_rho} compare as badly to the CDM predictions as they do to the galaxy fittings from \citep{Barranco:2018gjg}. Finally, we see that the surface gravity of the halos is extremely low, of about $10^{-11}- 10^{-13}$ cm/s$^2$. 
Then, we conclude, the halo system is not relativistic and the  $\frac{2G m}{c^2r}$ factor, that appears in the TOV equation (\ref{TOVpressure}), which determines the effects of General Relativity, is negligible. This is the reason why the Newtonian approximation produces the same results as the relativistic one when describing the dark matter halos in elliptical galaxies.

\section{Conclusions}\label{section_5}
 
As a model of the galactic halos we have considered very small, neutral DM particles,  which interact only gravitational with themselves and the baryonic matter. DM particle velocity dispersion acts as an intrinsic halo pressure which provides stability in the hydrodynamic analogy. The rotation curves of spiral galaxies determine completely the corresponding Newtonian gravitational potential $\Phi(r)$ of the static, spherically symmetric spacetime metric. This implies that the gravitational field inside the halo is weak, of the order of $\beta^2$, and the temporal term of the metric is not dependent on the equation of state used to describe the dark matter component.
Once a phenomenological rotational profile is proposed for a galaxy, if the dark matter is modeled as a perfect fluid, then the equation of state for dark matter is determined by the rotational velocity profile.
In particular, if the rotational velocity profile predicts a flat rotation curve for large radii, as the velocity profile proposed in \cite{Persic1996}, then the resulting EoS for dark matter,  given by equation \ref{EOS_rho}, has a barotropic limit for $\bar r \gg 1$. The resulting density profile decays as the inverse square power of the radial coordinate and in consequence, the density and pressure never reach zero at the boundary of the halo. Moreover, as was mentioned in Section II, the halo mass is divergent, increasing linearly in $r$. 
This fact implies that the invisible matter distributed in a spherical halo around spiral galaxies extends to infinity. Despite this, by defining the radius of the halo as the point where the halo density matches the average density of the Universe we can circumvent this problem. More troublesome, once the configurations that fit the observed rotational velocity curves of Low Surface Brightness Galaxies are found, they are in the unstable branch of possible configurations. 
This shows that although it is possible to find halos that fit the observed data, those configurations demand central pressures and densities that are too high regarding hydrostatic equilibrium, and therefore will be unstable under small radial perturbations. We conclude that if the Universal Velocity profile should be valid to fit the rotational velocity curve of galaxies and dark matter could be modeled as a perfect fluid, then one of our assumptions, i.e. hydrostatic equilibrium, spherical symmetry, or staticity of the spacetime, should not be valid, because with these assumptions the EOS is unique unless other microscopic parameters come into play, like temperature or density-dependent self-interactions.
Let us recall that the specific equation of state we have studied in this work has been fixed such as the resulting velocity profile for test particles will be given by the PSS profile. In other words, the EoS under study has a high degree of fine-tuning. On the other hand, it is known that dark matter with pressure modeled as a perfect fluid with microscopical physical EoS gives velocity profiles that decay at infinity. Nevertheless, that decaying velocity profiles can fit statistically observational rotational curves with central densities of selfgravitating DM halos in the stable branch \citep{Barranco:2018gjg}. Thus our methodology employed here can be seen as a test for proposed rotational velocity profiles. With this in mind, we envision two lines of work that could alleviate the problems discussed without modifying the spirit of the present work. The first is to consider the EOS obtained from a different density profile, for example, the Burkert profile. It seems that part of the problem is that the EOS that we consider is too soft, corresponding to a barotropic limit for low density, while the Burkert profiles corresponds to a polytropic limit with exponent approximately $1.2$. The second option is to try to include the temperature as part of the considered microscopic variables, although this entails making assumptions about the thermodynamic properties of DM.

\section*{Acknowledgments}

AA acknowledges the support of CONICET, Argentina, through grant PIP 112-201301-00532, and of Universidad Nacional de Cuyo, Argentina, through SIIP grant M060. JB and AB are partially supported by Conacyt-SNI. AB acknowledges support through Conacyt Ciencia de Frontera grant 304001.







\label{lastpage}
\end{document}